\documentclass[useAMS,usenatbib]{mn2e}

\title[Lithium in M4]
  {Lithium abundance in the globular cluster M4:\\
   from the Turn-Off to the RGB Bump
  \thanks{Based on observations collected at the ESO-VLT under program 081.D-0356. 
  Also based on observations collected at the ESO-MPI Telescope under program 69.D-0582, 
  and with the NASA/ESA HST, obtained at the Space Telescope Science Institute, 
  which is operated by AURA, Inc., under NASA contract NAS5-26555.}}
   
\author[A. Mucciarelli et al.]
  {A.~Mucciarelli,$^1$
   M.~Salaris,$^2$
   L.~Lovisi,$^1$
   F.R.~Ferraro,$^1$
   B.~Lanzoni,$^1$ 
   \newauthor
   S.~Lucatello,$^{3,4}$ 
   and R.G.~Gratton$^3$\\
  $^1$Dipartimento di Astronomia, Universit\`a 
  degli Studi di Bologna, Via Ranzani, 1 - 40127
  Bologna, ITALY
  \\
  $^2$Astrophysics Research Institute, Liverpool John Moores University, 
    12 Quays House, Birkenhead, CH41 1LD, United Kingdom  
   \\
   $^3$INAF- Osservatorio Astronomico di Padova,
   Vicolo dell'Osservatorio 5, I--35122 Padova, Italy
   \\
   $^4$Excellence Cluster Universe, Technische Universit¨at M¨unchen, 
    Boltzmannstr. 2, 85748, Garching, Germany
    }

\usepackage{graphicx}

\usepackage{color}
\pagerange{\pageref{firstpage}--\pageref{lastpage}} \pubyear{2002}

\def\LaTeX{L\kern-.36em\raise.3ex\hbox{a}\kern-.15em
    T\kern-.1667em\lower.7ex\hbox{E}\kern-.125emX}

\begin{document}

\label{firstpage}

\maketitle

\begin{abstract}
We present Li and Fe abundances for 87 stars in the globular cluster M4, 
obtained by using high-resolution spectra collected with GIRAFFE@VLT. The targets 
range from the Turn-Off up to the Red Giant Branch Bump. The Li abundance 
in the Turn-Off stars is uniform, with an average value equal to A(Li)=~2.30$\pm$0.02 dex 
($\sigma$=~0.10 dex), 
consistent with the upper envelope of Li content measured in other globular clusters 
and in the Halo field stars, confirming also for M4 the discrepancy with the 
primordial Li abundance predicted by WMAP+BBNS. 
The global behaviour of A(Li) as a function of the effective temperature allows to 
identify the 2 main drops in the Li evolution due to the {\sl First Dredge-Up} and 
to the extra-mixing episode after the {\sl Red Giant Branch Bump}. 
The measured iron content of M4 results to be [Fe/H]=~--1.10$\pm$0.01 dex 
($\sigma$=~0.07 dex), with no systematic offsets between dwarf and giant stars. \\
The behaviour of the Li and Fe abundance along the entire evolutionary path is incompatible with 
theoretical models including pure atomic diffusion, pointing out that an additional 
turbulent mixing below the convective region needs to be taken into account,
able to inhibit the atomic diffusion.
The measured value of A(Li) and its homogeneity in the Turn-Off stars allow to put strong 
constraints on the shape of the Li profile inside the M4 Turn-Off stars. 
The global behaviour of A(Li) with the effective temperature 
can be reproduced with different pristine Li abundances, depending on the 
kind of adopted turbulent mixing. One cannot reproduce the global trend 
starting from the WMAP+BBNS A(Li) and adopting the turbulent mixing described 
by \citet{richard05} with the same efficiency used by \citet{korn06} to explain the Li content 
in NGC~6397. In fact such a solution is not able to well reproduce simultaneously 
the Li abundance observed in Turn-Off and Red Giant Branch stars. 
Otherwise, the WMAP+BBNS A(Li) can be reproduced assuming a  more efficient turbulent mixing 
able to reach  deeper stellar regions where the Li is burned.\\ 
We conclude that the cosmological Li discrepancy 
cannot be easily solved with the present, poor understanding of the turbulence in the 
stellar interiors and a future effort to well understand the true nature of this 
non-canonical process is needed.
\end{abstract}


\section{Introduction}

The determination of the initial Li abundance in low-mass, metal-poor, 
Population II stars is one of the most debated 
astrophysical topics. Li is one of the few elements synthesized  
during the primordial Big Bang nucleosynthesis (BBNS), and its initial 
abundance in metal poor stars sets constraints to the BBNS. 

Li is a very fragile element, that is destroyed when 
the temperature is larger than $\sim$2.5 $10^6$ K through the 
$^{7}$Li(p,$\alpha$)$^{4}$He reaction.
Hence, whenever Li is produced in the stellar interiors during hydrostatic 
burnings, it is immediately disintegrated. 
The discovery of a constant Li abundance 
in unevolved, Population~II stars, 
the so-called {\sl Spite Plateau} \citep{spite82} --  has been interpreted
as the signature of the primordial abundance of Li, produced during the BBNS. 
When the lithium abundance is expressed as A(Li)=log(n(Li)/n(H))+12, 
the {\sl Spite Plateau} turns out to be between 2.1 and 2.4, 
depending on the adopted $T_{eff}$ scale 
\citep{boni97, charb05, asplund06, boni07, aoki09}.

The recent estimate of the cosmological baryon density obtained with the 
{\sl Wilkinson Microwave Anisotropy Probe} (WMAP) satellite 
\citep{spergel07} from the power spectrum of the cosmic 
microwave background fluctuations, throws into a crisis the classical 
interpretation of the {\sl Spite Plateau}. 
In fact, by combining the WMAP results with standard BBNS calculations, 
the most recent estimate of Li is 
a primordial abundance A(Li)=~2.72$\pm$0.06 dex \citep{cyburt08}
\footnote{Adopting different rates for the 
$^{3}$He($\alpha$,$\gamma$)$^7$Li reaction, \citet{steigman07} 
derived a slightly lower value, A(Li)=2.65, still in 
disagreement with the {\sl Spite Plateau}.}.
This value is significantly 
higher, by at least a factor of 3, than the Li abundance derived 
from metal-poor dwarfs.

At present, the WMAP+BBNS/{\sl Spite Plateau} discrepancy is 
still unsolved and different possible solutions have been 
advanced and investigated:
(1) an inadequacy of the standard BBNS treatment;
(2) Population III stars could have been capable to destroy some of the 
pristine Li and Population II stars were born 
from a Li-depleted gas \citep{piau06} ; 
(3) Li may be depleted in the photospheres of Population II stars 
(born with WMAP+BBNS Li abundance) by atomic diffusion: 
this process alters the surface chemical composition 
of the star, hiding a fraction of Li (but also He, Fe and other 
heavy elements) below the observable photospheric layers.

Galactic globular clusters (GCs) provide an alternative way to investigate 
the primordial Li abundance, for they host samples of stars 
approximately coeval, with uniform initial abundances of several 
elements -- Fe among them --  and whose evolutionary status  
is easily determined from photometry.  Also, GCs offer the possibility 
to trace the diffusion effects along the entire evolutionary path
of Population II stars. In fact, the amount of diffusion can be easily 
measured from the offset of the surface abundance of elements like Fe -- that are not 
involved in nuclear burnings --   
between the turn off (TO) stars
(where the diffusion affects appreciably the photospheric abundances) 
and the red giants (where convection restores the original photospheric abundances).
Systematic differences between the Fe content of Sub Giant Branch (SGB) and red giant 
branch (RGB) stars have been detected in M92 \citep{king98} 
and in NGC~6397 \citep{lind09b}. On the other hand independent studies have 
found a high level of homogeneity in Fe and other elements 
in the latter cluster \citep{gratton01} and in others 
\citep[see also][]{castilho00,cohen05}.

Extensive chemical screenings of Li content in GCs dwarf stars have been until 
now limited to the few closest objects,
namely 47 Tuc \citep{boni07,dorazi10}, NGC 6397 
\citep{boni02a, korn06, gh09, lind09b}, NGC 6752 \citep{pasquini05}
and M92 \citep{boes98, boni02b}. 
The metal-poor GC NGC 6397, in particular, has been recently 
the object of an intense debate regarding the discrepancy between 
cosmological and stellar Li abundance. \citet{korn06} and 
\citet{lind09b} find a trend between A(Li) and $T_{eff}$ along the SGB and 
the early RGB, roughly consistent with models \citep[][hereafter R05]{richard05} 
that start with the 
BBNS A(Li) and include atomic diffusion moderated by 
some ad-hoc turbulent mixing that brings back to the surface 
some of the Li (and other elements) diffused 
below the convective envelope during the main sequence (MS) phase. 
\citet{gh09} have studied a sample of MS and SGB stars in the same cluster, 
and found A(Li) decreasing with decreasing $T_{eff}$ in both groups of stars, 
with the SGB stars showing systematically higher values of A(Li) at a given $T_{eff}$. 
This pattern cannot be reproduced satisfactorily by any existing theoretical model
starting with the BBNS value of A(Li).

It is relevant to highlight that the investigation of A(Li) in 
GCs needs some special care, because of the occurrence of 
chemical anomalies in the light elements. In fact, all the GCs 
studied so far exhibit large star-to-star variations of C, N, O, 
Na, Mg and Al and in particular anti-correlations between O and Na, 
and between Mg and Al
\citep{carretta09a, carretta09b}. The widely accepted scenario to explain these 
patterns is that a second stellar generation was born in the cluster 
from the hydrogen-processed material ejected by the first population 
\citep{gratton01}. 
In this framework, Li could have a relevant role. A crucial point to 
recall is that the NeNa cycle, responsible for the pollution of the 
cluster gas, occurs at temperatures of 4--5$\cdot10^7$ K, 20--30 times higher 
than the Li-burning temperature. Thus, unless the polluting stars are also 
able to produce Li, the second generation stars should be 
Li-poor with respect to the first one and Li-O correlations and Li-Na 
anticorrelations are expected. 
Previous investigations about this issue lead to conflicting conclusions: 
\citet{pasquini05} observed for the first time the Li-Na anticorrelation 
(and also the Li-O correlation) in 9 TO stars of NGC~6752 and a similar 
behaviour has been found also by \citet{boni07} in 4 TO stars of 47 Tuc.
In their sample of NGC~6397 TO stars, \citet{lind09b} detected 3 stars 
out of 100 with low Li and high Na abundances, suggesting a possible 
Li-Na anticorrelation, while all the other TO stars share the same 
Li abundance over a range of $\sim$0.8 dex in A(Na). 
Recently, \citet{dorazi10} and \citet[][hereafter DM10]{dm10} presented Li abundances 
in TO stars of 47 Tuc and RGB stars of M4, respectively. In both cases, 
first and second generation stars share the same Li abundance, regardless 
of the Na and O content, but with different star-to-star scatter between 
the two stellar groups.

In this paper, we present the Li abundance measured from 
high-resolution spectra of a sample of stars ranging from the TO to the RGB Bump 
in the GC M4. 
This is the first determination of Li content in the dwarf stars of this cluster.
The Li content in the TO stars of M4 is then compared with the Li abundances 
measured in other GCs, in the field stars, and with the WMAP+BBNS 
predictions. The behaviour of A(Li) as a function of [Fe/H] is 
discussed in the light of stellar evolution models with and without 
atomic diffusion.
Also, the Fe content along the entire evolutionary path 
is discussed and it is used to quantify the efficiency of atomic diffusion.
Finally, we discuss the possible role of turbulent mixing in the 
theoretical models and the open questions related to the 
determination of the pristine Li abundance of M4.

\section{Observational Data}

\begin{figure}
\includegraphics[width=84mm]{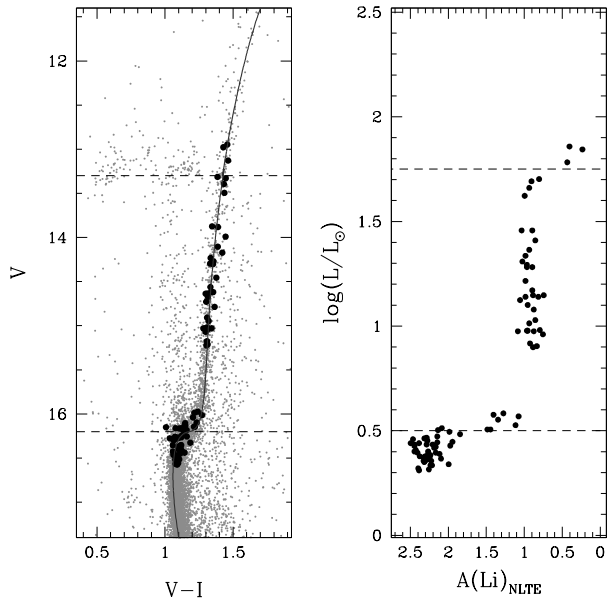}
\caption{Left panel: colour-magnitude-diagram of M4 (gray dots) 
with spectroscopic targets marked as black filled circles. The solid 
line is the best-fit isochrone (see text for details). Right panel: 
behaviour of A(Li) as a function of luminosity. The two
dashed horizontal lines denote the position of the two detected 
drops of A(Li) (see text for details).}
\label{fig1}
 
\end{figure}

This work is based on high-resolution spectroscopic observations 
of 87 stars in M4 performed with the GIRAFFE@FLAMES spectrograph 
mounted at VLT@ESO, 
within a project aimed to search for anomalous abundances in the 
BSS atmospheres \citep[see][]{ferraro06}.
All stars have been observed with the 
HR15N, HR18 and HR22 gratings that include the Li doublet 
at 6707.8 $\mathring{A}$, the $H_{\alpha}$ Balmer line and several Fe I lines.
The spectral sample includes 51 TO/SGB stars and 36 giants  
up to the RGB Bump magnitude level. Target selection, 
reduction of the spectra and measurement of the radial velocities 
are discussed in \citet{lovisi10}. The position of the 
targets in the (V, V-I) colour-Magnitude Diagram (CMD) 
is shown in Fig.~\ref{fig1} (left panel). The Signal-to-Noise Ratio 
(SNR) around the Li line is 
$\sim$50-80 for dwarf stars and rises up to about 300 for the 
brightest RGB stars. The membership of all the targets has been 
confirmed by the combined information of proper motions, radial 
velocities and metallicities, as discussed in \citet{lovisi10}.
Only one target in our sample has been found to be in common 
with the list of binaries by \citet{sommariva09} and it was excluded 
from the analysis.

\section{Analysis}

\subsection{Atmospherical parameters for TO/SGB stars}

Atmospherical parameters for SGB stars have been already presented 
in \citet{lovisi10}. Here we briefly discuss in more details only 
the derivation of $T_{eff}$.
Temperatures have been computed from a $\chi^2$ minimization between the 
observed wings of the $H_{\alpha}$ Balmer line and a grid of 
synthetic spectra. Synthetic spectra have been computed by means of 
the SYNTHE code from a grid of 1D LTE plane-parallel model 
atmospheres, obtained with ATLAS9 code in its Linux version
\citep{kur93a, kur93b, sbordone04}. Adopted model atmospheres are based on the 
NEW Opacity Distribution Functions described by \citet{castelli03} 
without inclusion of the {\sl approximate overshooting} and with 
$\alpha$-enhancement abundance patterns of [$\alpha$/Fe]=~+0.4 dex.
For the modeling of the Balmer line, we adopted the self-resonance 
broadening theory by \citet{ali65, ali66}.

The dominant source of error for $T_{eff}$ derived from the Balmer line 
is due to the residuals in flat-fielding and blaze function correction, 
that can introduce residual and spurious curvatures in the spectrum, 
afflicting the continuum placement. Such an effect is not easy 
to model, but in the fiber-fed GIRAFFE spectra it is less critical 
than in the slit spectra, since the flat-field is acquired through 
the same optical path used for the stellar spectra. 
In order to minimize this uncertainty, observed and synthetic spectra 
have been normalized by using the same continuum 
regions, typically $\pm$100 $\mathring{A}$ from the line centre.

In order to assess the internal error in the $T_{eff}$, 
we resorted to Monte Carlo simulations, similar to the procedures 
already adopted by \citet{boni07} and \citet{lind08}. 
The effect of the photon noise is quite small: we performed Monte Carlo 
simulations of 1000 synthetic spectra around the $H_{\alpha}$ line 
injecting Poissonian noise in order to reproduce the typical SNR of our 
spectra and we repeated the procedure to derive $T_{eff}$. We find 
that at SNR=~50 the error is of about 20 K, due to the large number of 
pixels used in the $\chi^2$ minimization. A similar check has been 
performed to assess the impact of the adopted fitting windows. 
Finally, we attribute to our $T_{eff}$ random errors of the order of 
30-40 K.
Other sources of uncertainties can affect the Balmer $T_{eff}$ and 
considered as systematic errors, for instance the gravity dependence 
of $H_{\alpha}$, the residual curvature in the flat-fielding, 
the adopted broadening theory and finite pixel-step in our spectra 
\citep[see e.g.][for a detailed discussion of these source of errors]{boni07}. 
In particular, 
\citet{boni07} and \citet{sbordone10} discussed the  
sensitivity of the $H_{\alpha}$ line to the gravity and found that it
is important in the low metallicity regime 
([Fe/H]$<$--2.0 dex), significantly lower than that of M4. 
For the TO stars we estimate 
a variation of only $\sim$10-20 K in $T_{eff}$ for a variation of 0.5 in log g, 
while for the coldest SGB stars such a sensitivity rises to $\sim$40 K. 
Bearing in mind these sources of uncertainty and  we conservatively 
attribute a typical, absolute error of 100 K to TO/SGB stars and of 150 K 
to the coldest SGB stars.

\subsection{Atmospherical parameters for RGB stars}

For RGB stars $T_{eff}$ cannot be inferred with the procedure described 
above, since, the $H_{\alpha}$ wings are not sensitive enough to $T_{eff}$ 
variations in cold stars. Some authors use the wings of $H_{\alpha}$
line to derive $T_{eff}$ also for RGB stars, but it is worth noticing that 
this method, when applied to giant stars, is subject to very high 
uncertainties (of 300 K at least). For stars with V$<$15.3, $T_{eff}$ 
have been derived by projecting their position 
in the CMD on the  best-fit isochrone.
(left panel of Fig.~\ref{fig1}, where the CMD is 
corrected for differential reddening). 
On the basis of the standard procedure \citep[see for example][]{piotto99}, 
we estimate that the differential reddening is of
the order of $\delta$E(B-V)=~0.25 in this cluster. 
The isochrone with 
appropriate metallicity (Z=~0.004, corresponding to [Fe/H]=$-$1.01 
for [$\alpha$/Fe]=+0.4) and age (12 Gyr), has been selected 
from the BaSTI database \footnote{http://albione.oa-teramo.inaf.it/} \citep{pietrinferni06}, 
and transformed into the observed plane by adopting E(B-V)=~0.32 and 
$(m-M)_0$=~11.30, consistent with the results by \citet{bedin09}. 
\footnote{The adopted isochrone is calculated without atomic diffusion. 
Including atomic diffusion would change the age of the isochrone of about 1 Gyr 
but the $T_{eff}$-colour relation would be essentially unchanged.}
It is worth noticing that because of the high value of the extinction, 
we applied a reddening correction that takes into account 
the dependence of $A_{\lambda}$/$A_V$ on the star surface gravity, $T_{eff}$ and [Fe/H]. 
We did this by following the same procedure described in \citet{bedin09}, adopting 
the extinction law of \citet{cardelli89}, re-scaled to $R_V$=~3.8, which is the value 
measured in the direction of M4 \citep{hansen04, bedin09}.

The colour-$T_{eff}$ transformations of the BaSTI database have been 
computed from the same model atmospheres adopted in our spectroscopic analysis. 
Hence such a $T_{eff}$ scale 
is formally homogeneous with that adopted for the dwarf stars. 
Indeed the BaSTI isochrone $T_{eff}$ scale 
well agrees with that derived from the $H_{\alpha}$ line for the 
TO/SGB stars: we find an average difference 
$T_{eff}^{BaSTI}$-$T_{eff}^{H_{\alpha}}=-$28~K 
($\sigma$=126 K) for stars with V$>$15.3. Thus, no relevant systematic 
offset is introduced between the abundances computed for dwarf and giant 
stars (the effect of different $T_{eff}$ scales on the derived
abundances is discussed in Sect. \ref{tscale}). Taking into account 
the photometric error, the uncertainty in E(B-V) and that 
arising from the differential reddening correction, we attribute 
a typical error of 100 K to $T_{eff}$ of RGB stars.

Similarly to the dwarf stars \citep[see][]{lovisi10}, the gravities have 
been obtained from the position of the stars in the CMD, and the
microturbulent velocities have been derived spectroscopically, 
by erasing any trend between A(Fe) and the equivalent width (EW). Main parameters, 
Li and Fe abundances for all the target are listed in 
Table~1, the complete version being available in 
electronic form.

\begin{table*}
\begin{minipage}{120mm}
\caption{Identification numbers, coordinates, temperatures, gravities, 
EW of the Li line, A(Li) and [Fe/H] abundances, SNR around the Li line for 
the observed stars. A complete version of the table is available in electronic form.} \label{anymode}
\begin{tabular}{lccllllllr}
\hline
ID & RA & Dec & $T_{eff}$ & log~g & EW(Li) &  A(Li)  & [Fe/H]  & SNR\\
\hline
   & (J2000)   &   (J2000)   &  (K)   &  (dex)   &  (m$\mathring{A}$)  &  (dex)  &  (dex)  & \\
\hline   
  8460 &  245.9471429	&  -26.3749119     &   5000   &   2.8    &   22.7    &   0.96     & -1.14      &     104     \\
  8777 &  245.9373405	&  -26.3613910     &   5150   &   3.2    &   11.8    &   0.79     & -1.13      &      85     \\
  9156 &  245.9069486	&  -26.3438488     &   4950   &   2.7    &   20.9    &   0.85     & -1.06      &     217     \\
 13282 &  245.7901731	&  -26.3909501     &   5850   &   3.9    &   39.5    &   2.11     & -1.00      &      53     \\
 28007 &  245.8049983	&  -26.6687472     &   5100   &   3.0    &   13.6    &   0.81     & -1.16      &     116     \\
 28890 &  245.8083640	&  -26.6339178     &   5850   &   3.8    &   30.5    &   1.98     & -1.06      &      62     \\
 28999 &  245.7440673	&  -26.6298102     &   6000   &   4.0    &   34.4    &   2.16     & -1.20      &      68     \\
 29074 &  245.8010945	&  -26.6273020     &   6150   &   3.9    &   34.8    &   2.28     & -1.00      &      48     \\
 29411 &  245.7990049	&  -26.6149585     &   5500   &   3.7    &   15.9    &   1.34     & -1.05      &      60     \\
 29643 &  245.7572145	&  -26.6058864     &   5800   &   3.8    &   40.3    &   2.08     & -1.19      &      68     \\
 29729 &  245.7547183	&  -26.6022333     &   5950   &   3.9    &   46.8    &   2.28     & -1.18      &      58     \\
 30167 &  245.7958060	&  -26.5866438     &   5950   &   3.8    &   35.6    &   2.14     & -1.04      &      60     \\
 30475 &  245.7195282	&  -26.5759641     &   5100   &   3.0    &   12.0    &   0.74     & -1.05      &     158     \\
 30922 &  245.7432036	&  -26.5602858     &   6050   &   4.0    &   33.1    &   2.18     & -1.16      &      48     \\
\hline
\end{tabular}
\end{minipage}
\end{table*}

\subsection{Lithium}

Li abundances have been measured by comparing the observed 
EWs of the Li I resonance doublet at 6707.8 $\mathring{A}$ 
with the EW computed from a grid of synthetic spectra. 
A curve of growth has been built for each star and we derived 
A(Li) interpolating each curve at the observed EW value. 
Due to the small wavelength separation of the Li components, 
this line can be reasonably approximated with a single gaussian 
profile at the GIRAFFE resolution. EWs have been measured with our 
own FORTRAN procedure, that performs a gaussian fit on a spectral 
window selected interactively. The continuum level is traced locally 
by considering a region of $\sim$20 $\mathring{A}$ around the Li line,  
rejecting spikes, cosmic rays and spectral features with a 
$\sigma$-clipping algorithm and adopting as continuum level the peak 
of the flux distribution computed from the surviving points. 
As sanity check we performed an independent EW measurement by using 
the IRAF task {\sl splot}, finding an average difference 
$EW_{IRAF}$--$EW_{fit}$=~--0.09 m$\mathring{A}$ with a dispersion 
$\sigma$=~2.35 m$\mathring{A}$.
Such a small difference is irrelevant to the purposes of this work.

Atomic data for the Li doublet, including its hyperfine structure 
and isotopic splitting (with $N(^{6}Li)/N(^{7}Li)$=~8\%), 
are from \citet{yan95}. Corrections for NLTE 
effects have been applied by interpolating the grid of 
\citet{carlsson94}: derived NLTE corrections are quite small, of the 
order of 0.01 dex
\footnote{Note that the adoption of the NLTE corrections of 
\citet{lind09a} would have a negligible (of about a few hundredths) 
impact on the Li abundances.}.

The error in A(Li) is computed by adding in quadrature the two main sources 
of uncertainty, namely the errors arising from $T_{eff}$ and from 
the EW measurement. A(Li) turns out to be very sensitive to the 
used $T_{eff}$, because Li is almost totally ionized. The typical error 
in $T_{eff}$ ($\pm$100 K) translates in an error in A(Li) of $\pm$0.07 dex 
for TO/SGB stars and slightly higher ($\pm$0.09 dex) for RGB stars. 
It is customary to estimate the error in EW measurement by using the 
Cayrel formula \citep{cayrel88}. This gives 
$\pm$4-4.5 m$\mathring{A}$ for TO/SGB stars and $\pm$1-2 m$\mathring{A}$ 
for RGB stars, translating in A(Li) errors of the order of 
$\pm$0.05-0.07 dex and of $\pm$0.01-0.02 dex, respectively. 
The Cayrel formula neglects the uncertainty due to the continuum 
location, that can have a relevant impact in spectra with low SNR. 
We estimated the continuum location uncertainty from the average spread
of the residual spectrum around the Li line and performing several tests 
with different assumptions of the continuum level. Finally we derive  
an uncertainty of $\pm$0.10 dex for the TO/SGB stars with lowest SNR and a few 
hundredths of dex for the RGB stars.
Other sources of errors (linked to gravity and microturbulent velocity) 
can be assumed to be negligible since they correspond to an error in A(Li) 
of $\pm$0.01 dex each. 
Final absolute errors in A(Li) range between $\pm$0.08 and $\pm$0.15 dex, 
according to the stellar evolutionary stage.

\subsection{Iron}

Our spectral dataset allows to measure the Fe abundance 
in both TO/SGB and RGB stars.
In turn these measures can be used as tracers of the occurrence of 
atomic diffusion processes along the evolutionary path. 
Fe abundances for all the target stars have been derived 
from the EW measurement by using WIDTH code
\footnote{We have employed a version of the original code by R. L. Kurucz 
modified by F. Castelli, in order to use van der Waals damping constants 
by \citet{barklem00} when available. The source code is available 
at http://wwwuser.oat.ts.astro.it/castelli/sources/width9.html.} 
in its Linux version \citep{sbordone04}.
We employed two different linelists for TO/SGB and RGB stars, in order to 
include only unblended transitions, by taking into account the different 
degree of blending and line strength between dwarf and giant stars.
The number of Fe I lines is limited by the relatively small wavelength 
coverage of our spectra (we remark that the spectral regions sampled 
by the gratings HR18 and HR22 are strongly affected by deep 
telluric absorptions). 
Finally, we derived the abundance from $\sim$6-14 and $\sim$18-22 
Fe I lines, for TO/SGB and RGB stars respectively. Atomic data are from 
the classical line compilation by \citet{fuhr88}. 
As solar reference we adopted $A(Fe)_{\odot}$=~7.5 dex \citep{gs98}. 
The error estimate has been performed similarly to what we did for Li.
Taking into account the typical uncertainty on the atmospherical parameters we 
estimate a systematic error of about $\pm$0.10 and $\pm$0.15 dex for SGB and RGB stars, 
respectively. Random error obtained from the Cayrel formula and the 
continuum level uncertainty 
ranges from $\pm$0.14 (for SGB) to $\pm$0.03 dex (for RGB stars).

\section{Evolution of A(Li)}

\begin{figure}
\includegraphics[width=84mm]{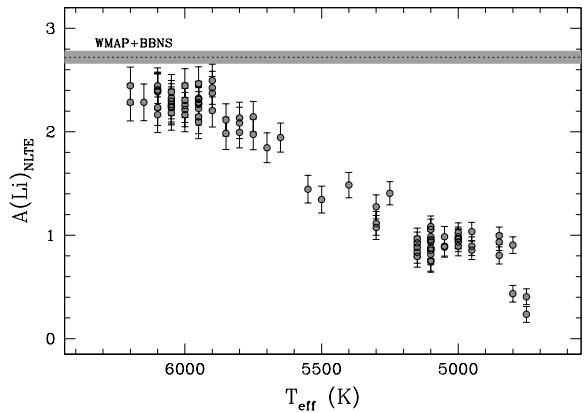}
\caption{Behaviour of A(Li) as a function of $T_{eff}$ for 
individual stars. The error bars include both random 
and systematic errors. The dotted line denotes the WMAP+BBNS value of  
A(Li) computed by \citet{cyburt08} and the grey area the 
corresponding uncertainty.}
\label{fig2}
\end{figure}

The right panel of Fig.~\ref{fig1} shows the behaviour of A(Li) as a function 
of luminosity for the observed targets. Luminosities have been 
derived from the best-fit BaSTI isochrone discussed above and according 
with the position of the star along the evolutionary sequences. 
The two horizontal dashed lines denote the magnitude and luminosity levels 
where two abrupt drops of A(Li) occur. Fig.~\ref{fig2} shows the values  
of A(Li) as a function of $T_{eff}$. We can describe the evolutionary behaviour 
of the Li abundance by identifying four regimes in Fig.~\ref{fig1} 
and ~\ref{fig2}:
\begin{itemize}
\item{\bf TO} ($T_{eff}\ge$5900 K; lg(L/$L_{\odot})<$0.5): stars located 
at the TO level have an average $<$A(Li)$>$=~2.30$\pm$0.02 dex ($\sigma$=~0.10).
To check whether the distribution of Li abundances is consistent -- 
given the observational random errors -- with a uniform value $A(Li)$=2.30, we performed 
the following test \citep[see, e.g., an application of the same test 
to the analysis of initial He abundances in GCs and of GC ages by][]{cassisi03, chab96, salaris02}.
For each individual star we have calculated a set of synthetic Li abundances by randomly 
generating -- using a Monte Carlo procedure -- 10000 abundance values, according to a Gaussian distribution with mean value 
equal to $A(Li)$=2.30, and $\sigma$ equal to the empirical random abundance errors. 
As random errors we have considered the errors due to the EW measurements 
(as described in Sect.~3.3) and the random errors in the $T_{eff}$ determination (as described in Sect.~3.1).
This is repeated for all 35 TO stars, and the final 350000 synthetic values are combined to produce an "expected'' 
distribution for the entire sample, in the assumption that the detected Li abundance dispersion is not intrinsic, 
but due to the individual errors, assumed Gaussian. 
The statistical F-test was then applied to determine if this "expected'' 
distribution (well sampled by the extremely large number of elements), displays a variance that is statistically consistent 
with the observed distribution of 35 objects. As customary, we state that an intrinsic A(Li) range of values 
does exist if the probability P that the 
two distributions have different variance is larger than 95\%.  

We find a value of P below 90\%, and therefore we can conclude that the observed dispersion  
is formally consistent with the random errors  
associated to the Li estimates of each star, and that our measurements provide 
a uniform Li abundance $A(Li)$=2.30$\pm$0.02 for all objects with $T_{eff}\ge$5900~K

This result 
points toward a general homogeneity of the Li abundance in M4 stars, in contrast 
to the large star-to-star scatter observed in other clusters (as 47 Tuc and NGC 6752) that
indicates an intrinsic dispersion in A(Li). 
\item{\bf SGB} (5200 K $< T_{eff}<$5900 K; 0.5$<$lg(L/$L_{\odot})<$0.6):
these stars are located along the SGB, from the TO to the base of the RGB. 
We found a decrease of A(Li) for increasing $T_{eff}$, 
with A(Li) ranging from 2.0 (in the hot SGB stars) to 0.9~dex (at the end of the SGB).
\item{\bf lower-RGB} (4800 K $< T_{eff}<$5200 K; 0.9$<$lg(L/$L_{\odot})<$1.7): 
stars in this luminosity range, corresponding to the portion of the RGB fainter than 
the RGB Bump (that occurs for this age and metallicity at lg(L/$L_{\odot})$=1.73), 
share the same abundance of Li, defining a plateau 
$<$A(Li)$>$=~0.92$\pm$0.01 dex ($\sigma$=~0.08).
\item{\bf upper-RGB} ($T_{eff}<$4800 K --- lg(L/$L_{\odot})>$1.7):
these stars are brighter than the RGB Bump and display a sudden drop of A(Li), 
reaching A(Li)$\sim$0.23 dex.
\end{itemize}

We remark that M4 is the second GC studied so far, after NGC 6397 
\citep{lind09b}, where A(Li) has been measured along the 
TO-SGB-RGB sequence: in both clusters two drops of A(Li)  
have been clearly detected, consistent with 
the pattern already seen in Halo stars \citep{gratton00}.
These drops can be associated with two distinct mixing episodes:\\ 
{\sl (i)} the canonical mixing episode called {\sl First Dredge-Up} that occurs 
when the convective envelope reaches inner regions 
where Li has been destroyed through the nuclear reaction 
$^7$Li(p,$\alpha$)$^{4}$He; this produces a drop of A(Li) by a factor 
$\sim$ 15 along the SGB;\\ 
{\sl (ii)} the extra-mixing episode after the RGB Bump, that occurs when the 
outward moving H-burning shell reaches the discontinuity in the H-profile left over 
by the convective envelope at its maximum penetration 
\citep[see e.g.][]{charb}. 
The stars at V$\sim$13.5 
\citep[where the RGB Bump is located, see e.g.][]{ferraro99} share the same 
abundance of the fainter RGB stars, while the 3 RGB stars above this luminosity display a 
drastic decrease of A(Li). When the mean molecular weight barrier is removed 
after the RGB Bump, the surviving surface Li is destroyed and the stars 
exhibit A(Li) approaching zero.

\begin{figure}
\includegraphics[width=84mm]{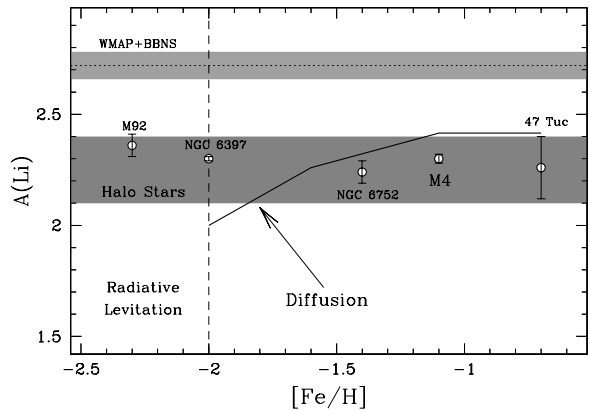}
\caption{A(Li) mean values for TO stars in the GCs studied so far, as a function of the 
initial cluster iron abundance measured from the RGB stars. 
Errorbars indicate the dispersion around the mean,  
normalized to the root mean square of the number of used stars. 
The dark grey area displays the range of values obtained for 
Halo stars by different recent investigations 
\citep{boni97, charb05, asplund06, boni07, aoki09}. 
The solid line displays the theoretical prediction for the value of A(Li) in TO 
stars for an age of 11-12~Gyr, in case atomic diffusion is fully efficient, starting 
from an initial WMAP+BBNS value A(Li)=2.72. The vertical dashed line shows the 
boundary for the radiative levitation 
to start affecting appreciably the photospheric chemical abundances of Li (see Sect.~8 for details).
The primordial A(Li) provided by WMAP+BBNS calculations and the associated uncertainty is 
also displayed (light grey region).}
\label{figfin}
\end{figure}

The Li abundance measured in M4 TO stars, A(Li)=2.30 dex, is in agreement 
with what found in all other GCs investigated so far
(Fig.~\ref{figfin}).
The metal-poor ([Fe/H]=$-$2.3 dex) GC M92 shows a mean value of 
A(Li)=2.36$\pm$0.05 dex, $\sigma$=~0.18, \citep{boni02b}. 
For NGC~6397 ([Fe/H]=$-$2.0 dex) different studies point toward A(Li) between 
2.25$\pm$0.01 \citep{lind09b} and 2.30$\pm$0.01 dex \citep{gh09} in stars along the MS 
or around the TO.
The value of A(Li) in 9 TO stars of NGC 6752 ([Fe/H]=$-$1.4 dex) has been measured by 
\citet{pasquini05} who found A(Li)=2.24$\pm$0.05 dex ($\sigma$=0.15). 
Finally, the metal-rich cluster 47 Tuc \citep{dorazi10} shows a similar mean 
abundance, but with a large star-to-star scatter, 
A(Li)=2.26$\pm$0.14 dex ($\sigma$=~1.38).

Even if small offsets can be present between these Li determinations
(because of the different temperature scales and adopted NLTE corrections), 
all these measurements are consistent each other and with the 
{\sl Spite Plateau} detected in the Halo dwarf stars (grey region). 
Hence, we confirm for M4 the Li content discrepancy with respect to the 
WMAP+BBNS results, already detected in the field stars and in other GCs.

\section{Evolution of [Fe/H]}

\begin{figure}
\includegraphics[width=84mm]{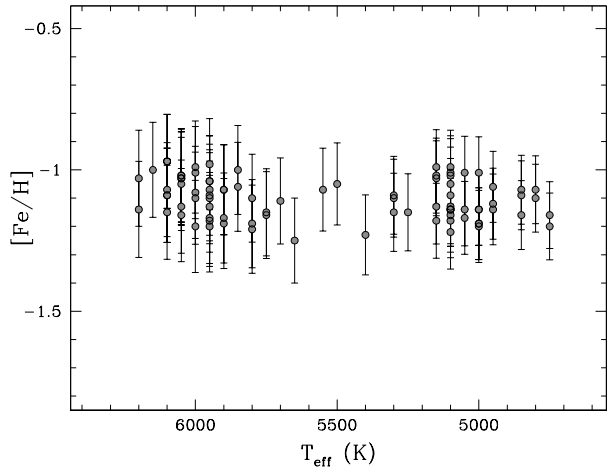}
\caption{Behaviour of [Fe/H] as a function of $T_{eff}$ for the individual stars 
in our sample. Errorbars include random and systematic errors.}
\label{fig3}
\end{figure}

Fig.~\ref{fig3} shows the estimated [Fe/H] as a function of $T_{eff}$ for the 
individual stars in our sample.
The iron abundance does not show any systematic trends with the temperature. 
By taking into account the entire sample, 
the average Fe abundance for M4 turns out to be [Fe/H]=$-1.10\pm0.01$~dex 
($\sigma$=0.07).  
This value basically agrees with previous spectroscopic analyses. 
\citet{ivans99} derived an average iron content [Fe/H]=~--1.18$\pm$0.01 dex,
\citet{marino08} obtained [Fe/H]=~--1.07$\pm$0.01 dex, 
\citet{carretta09c} found a value [Fe/H]=~-1.18$\pm$0.02 dex. Taking into 
account the adopted solar reference values, the absolute differences between 
our Fe determinations and the previous ones are of +0.06, --0.01 and +0.04 dex, 
respectively.
By dividing the 
sample in 3 groups, e.g. considering separately 35 TO stars, 16 SGB stars and 36 RGB stars, 
the derived average Fe abundances are [Fe/H]=$-1.08\pm$0.01 ($\sigma$=0.07), 
$-1.13\pm$0.03 ($\sigma$=0.07) and $-1.11\pm$0.01 ($\sigma$=0.07), 
respectively. Hence, no evident trend is detected between 
dwarf and giant stars.
A F-test analogous to the one described for the TO sample has been performed on the 
[Fe/H] values of our whole sample of M4 stars, and has provided -- as expected -- 
no statistically significant indication of an intrinsic spread.

Note that this analysis does not include any correction for the departures from 
LTE, due to the lack of NLTE corrections for individual Fe I lines at the 
metallicity of M4. Moreover, the precise order of magnitude of NLTE corrections 
for Fe lines is not well constrained, because some assumptions must be 
introduced (for instance the factor $S_H$ needed to correct the H I collisional 
cross-sections derived by the formula by \citet{drawin68} and the accuracy and completeness 
of the adopted model Fe atom). Although the departures from LTE are more 
critical at lower metallicities, we have tried to use the corrections computed 
by \citet{gratton99} for a representative high excitation Fe I transition 
\footnote{Note that the majority of the adopted Fe I lines in our analysis 
has excitation potential $>$ 3 eV.}, for an overall metallicity [M/H]=$-$1.0 dex. 
These NLTE--LTE corrections are positive and quite small ($\sim$0.05 dex for dwarfs and 
$\sim$0.03 dex for giant stars) and, if applied, they do not change 
the observed trend and are not capable to introduce a systematic offset 
in [Fe/H] between TO/SGB and RGB stars.

\section{Sanity checks}

\subsection{Median spectra}
To reduce the error sources and strengthen our result about the SGB stars 
(that have the lowest SNR), we averaged their spectra according to their 
individual $T_{eff}$. We obtained five spectra with enhanced SNR ($\sim$150 around 
the Li line) and repeated the entire analysis for each of them: normalization of
the spectrum, re-computation of $T_{eff}$ by fitting $H_{\alpha}$ wings and 
measurements of EWs for Li and Fe. The resulting new set of $T_{eff}$ well 
resembles the average of the single spectra values within $\sim$40 K, confirming 
the reliability of the individual $T_{eff}$. Also, A(Li) and [Fe/H] measured 
in these median spectra agree within few hundredths of dex with the average 
Li and Fe abundances computed from the individual stars.

\subsection{The impact of different $T_{eff}$ scales}
\label{tscale}

The Li abundance shows a plateau for temperatures larger than $T_{eff}$=~5900 K 
(see Fig.\ref{fig2}). The reliability of this trend is crucial, because 
stellar evolutionary models including pure atomic diffusion predict an 
increase of A(Li) before the dilution due to the {\sl First Dredge-Up} 
\citep[see e.g. the discussion in][]{salaris01}.

The most relevant source of error in the A(Li) determination is the assumption 
of the $T_{eff}$ scale. We therefore repeated the analysis by adopting 
two different photometric $T_{eff}$ scales, derived by means of the most recent 
$(V-I)_0$--$T_{eff}$ relations by \citet[][GHB09]{ghb09} and \citet[][C10]{casa10}, 
both based on the InfraRed Flux Method (IRFM). 
We remark that these $T_{eff}$ scales are independent of the $T_{eff}$ derived 
by the $H_{\alpha}$ line and affected by different source of errors. In particular, 
$T_{eff}$ inferred by IRFM are largely insensitive to uncertainties in model atmospheres 
(i.e. the adopted line broadening recipe, the assumption of 1-dimensional 
geometry) at variance with those derived from the $H_{\alpha}$ line, but they are 
affected by uncertainties on the photometry, absolute and differential 
reddening, photometric calibration and so on.\\ 
We have derived $(V-I)_0$ colours for our stars from the WFI photometry, corrected for differential 
reddening effects (see Sect. 3.2). By adopting the GHB09 calibration, we find an average 
difference $T_{eff}^{GHB09}$--$T_{eff}^{H_{\alpha}}$=$-$77~K ($\sigma$=~125 K). Similarly, 
the temperatures computed by the C10 calibration provide a mean difference 
 $T_{eff}^{C10}$--$T_{eff}^{H_{\alpha}}$=$-$29~K ($\sigma$=~123 K). 
These mean differences are small in absolute terms and also much smaller than the dispersion 
around their mean values. 
We therefore conclude that no significant difference is found between the $T_{eff}$ scales 
obtained from the Balmer line and from the IRFM. This finding reduces the major  
disagreement between previous generations of theoretical $(V-I)_0$--$T_{eff}$ transformations 
and IRFM calibrations \citep[see the extensive discussion by][]{weiss99}.
 
We have however re-analyzed our stars by employing also these two $T_{eff}$ scales, re-computing
both gravities and microturbulent velocities. The adoption 
of IRFM $T_{eff}$ scales does not change the global behaviour of our abundances, both Li and Fe. 
In particular, the iron content remains consistent between TO and RGB stars, 
without systematic offset between the different group of stars. 
Variations of about $\sim$-0.05 dex or less are obtained for the average 
A(Li) of TO stars.

These checks further reinforce the conclusion that the observed 
behaviour of A(Li) and [Fe/H] as a function of $T_{eff}$ is not an artifact  
due to an incorrect $T_{eff}$ scale.

\subsection{Comparison with \citet{dm10}}

\begin{figure}
\includegraphics[width=84mm]{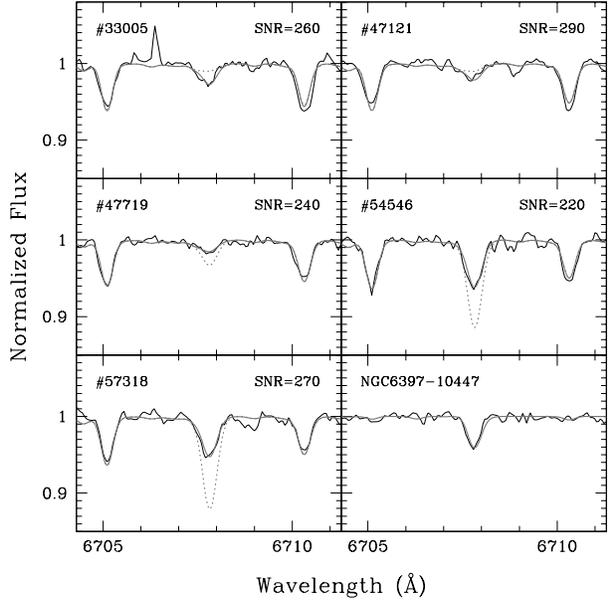}
\caption{Portions of the GIRAFFE spectra (solid black) around the Li line 
for 5 RGB stars in common with DM10. Solid grey lines show 
synthetic spectra computed with our Li abundances, and dotted grey lines 
display spectra computed with Li abundances from DM10. 
The bottom-right panel shows the Li line 
in the spectrum of star 10447 belonging to NGC~6397, from the dataset by \citet{lind09a}, 
compared to a synthetic spectrum computed with their atmospherical parameters 
and Li abundance A(Li)=1.04 (see text for details).}
\label{dmtest}
\end{figure}

The only other sample of Li abundance in M4 stars available to date 
concerns 104 RGB stars discussed by DM10. 
There is a small overlap between the 2 samples, with only 5 RGB stars in common. 
The agreement with their atmospherical parameters is 
satisfactory, despite different methodologies 
have been applied (i.e., for $T_{eff}$ we use the 
comparison with the BaSTI isochrone, while DM10 relies on the 
excitation equilibrium method). Our $T_{eff}$ values are higher by  
+36~K ($\sigma$=~22 K) on average, and also the gravities show very small differences 
(of the order of 0.2~dex). The mean difference between the 
iron abundances of these 5 targets is of only 
$-$0.04 dex ($\sigma$=~0.06 dex), including also a small offset in the solar zero-point. 
Although atmospherical parameters and iron abundances are fully consistent 
between the two studies, a large ($\sim$0.4 dex) difference is revealed 
in the Li content of 3 stars, while for the other 2 objects DM10 
provide A(Li) close to zero, lower than our measures.  
The discrepancies are not attributable to the quoted measurement errors,
nor to the difference in $T_{eff}$ (which leads to 
a difference in A(Li) of $\sim$0.02 dex only). 
Fig.~\ref{dmtest} shows the Li line in our spectra for the five stars in common,  
compared with synthetic spectra computed with our abundances 
(solid grey curve) and with those by DM10 (dotted grey curve). 
Note that for the 2 stars with A(Li)$\sim$0 in DM10, the 
differences are consistent with the different SNR of 
the spectra, because in their spectra (with typical SNR 50-100) such weak 
Li lines are hidden in the noise envelope. For the other 3 stars the discrepancy 
is more critical.

DM10 have employed ATLAS9 model atmospheres 
computed with the option of {\sl approximate overshooting}. 
As pointed out by \citet{molaro95} and \citet{bonifacio09}, the 
temperature stratification of ATLAS9 model atmospheres with and without 
{\sl approximate overshooting} can be very different in the optical depth 
range --1$<\lg(\tau)<$1, where the bulk of the weak/moderated lines 
is formed (including the Li line adopted in this work). 
We computed some ATLAS9 model atmospheres with 
{\sl approximate overshooting} and compared the synthetic Li line profile 
in the two cases. At the metallicity of M4, the difference between 
the thermal structures of the two models is less severe than at 
lower metallicities \citep[as those investigated by ][]{bonifacio09}, 
and the adoption of the ATLAS9 models with {\sl approximate overshooting}
leads to an increase of the Li abundance smaller than 0.1~dex.

As a sanity check, we apply our procedure to the GIRAFFE 
dataset of NGC~6397 stars studied by \citet{lind09b}. 
We retrieved the reduced data from the GIRAFFE Archive 
maintained at the Paris Observatory  
\footnote{http://giraffe-archive.obspm.fr/} and 
derived A(Li) for a sub-sample of their targets 
by using our procedure and their atmospherical parameters and 
EWs. Differences of a few hundredths of dex with respect to 
the derived A(Li) by \citet{lind09b} have been found.
The bottom-right panel of Fig.\ref{dmtest} shows a RGB star from the 
sample by \citet{lind09b}, in comparison with a synthetic spectra 
computed with their atmospherical parameters and A(Li).
From this test, we exclude any relevant systematic offset in our procedure 
to infer the Li abundance. 
As a further check, we redetermined the EWs of the Li line in 
a number of \citet{lind09b} stars, finding negligible differences. This confirms 
the reliability of our method to measure the EWs.
These two tests give further support to
the correctness of our A(Li) values for the RGB stars, but 
the discrepancy with DM10 remains unexplained and we 
postpone this issue to future investigations. 

Finally, we note that if our sample was affected by a systematic 
underestimate of $\sim$0.3-0.4 dex for both RGB and TO stars, 
we would obtain A(Li)$\sim$2.7 dex 
at the TO of M4, a value much higher than the A(Li) measurements along the 
{\sl Spite Plateau} and in general higher than all the Li measurements 
in Pop~II stars of the last three decades.

\section{Li abundance in first and second generation stars}
\label{first}

As already mentioned in Section 1, the current interpretation of the
anticorrelations observed in the GCs, involves a first generation of stars with 
efficient CNO, NeNa, and MgAl cycles in their interiors, whose chemically 
processed matter is transported to the surface, and ejected in the existing 
interstellar medium through mass loss processes. 
A second generation of stars is then formed out of this polluted matter, so 
that nowadays individual GCs are populated by a combination of first and second 
generation stars.  
The nature of this polluting first generation stars is still debated, with 
two competing candidates: intermediate-mass asymptotic giant branch (AGB) stars  
\citep{ventura09} or fast rotating massive stars \citep{decr07}. 
In both cases Li is expected to be depleted, since the NeNa cycle occurs at  
temperatures higher than that of Li burning. However, 
if AGB stars are responsible for intracluster pollution, they may have also produced Li, 
through the Cameron-Fowler mechanism \citep{cameron71}, so that the surface A(Li) may be able 
to increase to values similar to the initial abundances.
Thus, it is crucial to compare the Li abundance in first and second generation 
stars, bearing in mind that only the former should be born from a gas with 
the cosmological A(Li).
To this purpose, we study the behaviour of A(Li) as a function of [O/Fe]. 
The [O/Fe] abundance ratios have been derived by \citet{lovisi10} 
by measuring the permitted O~I triplet at 7771-74 $\mathring{A}$. 
Unfortunately, our dataset does not include Na transitions, usually 
adopted to well separate the two populations \citep[see][hereafter DM10]{lind09b, dorazi10}, 
because the choice of the setups has been done for different purposes 
\citep[see][]{lovisi10}.

\begin{figure}
\includegraphics[width=85mm]{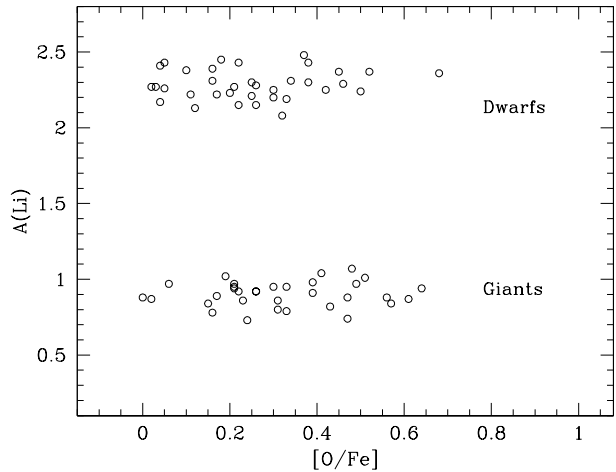}
\caption{Behaviour of A(Li) as a function of [O/Fe] for the stars of M4, 
excluding those located along the two Li drops (see Fig.~\ref{fig1}).}
\label{liox}
\end{figure}

Fig.~\ref{liox} shows the distribution of TO and RGB stars 
(excluding the stars located along the two Li drops, see Fig.~\ref{fig1}) in the 
[O/Fe]-A(Li) plane. For both dwarf and giant stars, the Li abundance 
does not show any correlation with [O/Fe], 
similar to what found in 47 Tuc \citep{dorazi10}
and in M4 by DM10. Measuring the Spearman rank coefficient, 
we find $C_S$=~0.05 and 0.06 for dwarf and giant stars, respectively. 
Such a test confirms the lack of correlation between Li and O content.

Hence, the Li abundance appears to be very similar among first and second generation 
stars. Adopting as boundary between the two populations the median value of the 
oxygen distribution ([O/Fe]$\sim$0.30 dex), we find very similar average A(Li) 
values, both for dwarf and giant stars. This finding points out that 
Li production occurs in this cluster.

In fact, if the progenitors of second generation stars are massive objects, they have 
destroyed their original Li content in their envelopes, whereas if AGB stars are 
responsible for intracluster pollution, they may have produced Li, through the 
Cameron-Fowler mechanism \citep{cameron71}, so that the surface A(Li) may be able 
to increase to values similar to the initial abundances. On the other hand, 
a point to recall is that the Cameron-Fowler mechanism to produce Li needs to
some fine-tuning. The entire envelope of the AGB stars 
where Li has been produced must be ejected before the newly formed Li 
is destroyed. 
The uniformity of the A(Li) values around the TO and along the lower RGB points 
somewhat to Li production in the polluters, as already discussed by DM10. 
In fact, since our sample of stars at a given $T_{eff}$ contains a random 
mixture of stars belonging to both first and second generations, if the polluters destroy 
Li we would expect an intrinsic spread in A(Li). Conversely, if the polluters produce Li, 
a quite fine tuned scenario is required to achieve a uniform 
Li abundance for both first and second generations of stars in M4. 
In any case, several uncertainties (as the mass-loss rates and the 
Li yields) dramatically affect these kind of models.

\section{Theoretical models}

To compare our measured Li and Fe abundances with the results from theoretical 
stellar evolution models -- with and without including the effect of atomic diffusion --  
we have employed the stellar evolution code described in 
\citet{pietrinferni06}. More in detail, we have considered an initial chemical 
composition with Y=~0.25, [Fe/H]=~$-1.1$, and the same $\alpha$-enhanced metal 
mixture as in \citet{pietrinferni06} and 
\citet{salaris01}, with an average [$\alpha$/Fe]=~+0.4 dex and 
with initial A(Li)=~2.72, corresponding to the WMAP+BBNS Li prediction. 
We have then computed the evolution of models with masses of 
the order of 0.85-0.89~$M_{\odot}$ and TO age equal to 12~Gyr 
(for models without diffusion) and 11~Gyr (for models with diffusion). 
These objects are assumed to be the counterpart of stars currently evolving 
along the SGB and RGB of the cluster.  The evolutionary sequences were computed 
starting from the pre-MS phase, until they reached $T_{eff}\sim$4700~K along 
the RGB, i.e. below the lower $T_{eff}$ limit of our observed stars. 

Diffusion was included by solving the Burgers equations 
for a multicomponent fluid following \citet{thoul}, and we considered 
explicitly the evolution of  H, He, Li, C, N, O, and Fe abundances. 
The other elements have been assumed to diffuse like Fe. 
Our calculations account for the effects of temperature, gravity and 
molecular weight gradients, while radiative levitation is not included. 

As for the computation of the radiative opacity, given that for this 
metallicity all heavy elements are approximately equally 
diffused, the abundance ratios are not significantly affected, and the effects 
on the stellar opacity are accounted by interpolating among 
$\alpha$-enhanced opacity tables of different Z. 
Superadiabatic convection is treated according to the mixing length 
formalism of \citet{cox}. The value of the mixing length has been 
chosen from a solar model calibration \citep[see e.g.][]{pietrinferni06}. 
Finally, the rates for the Li destruction reactions are taken from the 
NACRE compilation \citep{angulo99}.

We remark that the only element transport mechanisms accounted for in our models 
are convection (with boundaries fixed by the Schwarzschild criterion and no overshooting) and 
atomic diffusion. No additional mixing process able to operate in 
RGB stars brighter than the RGB bump, or turbulent 
mixing that moderate the effect of atomic diffusion are included.

Given that our analysis focuses on the theoretical interpretation of the observed 
Li and Fe abundances, it is important to assess whether neglecting the radiative 
levitation --  that tends to counteract the effect of temperature, gravity and 
molecular weight gradients -- can affect our conclusions.
Regarding the predicted surface Li abundances, Fig.~3 of \citet{richard02} and 
Figs~3 and 4 of \citet{korn06b} show that for metallicities above 
[Fe/H]$\sim -$2 the effect of neglecting radiative levitation is negligible.

As for the surface Fe abundances, the inclusion of radiative levitation at metallicities above 
[Fe/H]$\sim -$1.3 causes only small changes that do not affect any of the conclusions of our study. 
To give a quantitative example, we calculated a  
[Fe/H]=$-$1.61, 12~Gyr isochrone, and compared the surface [Fe/H] values  
with the corresponding predictions in Fig.~9  of \citet{richard02}, that include also 
the effect of radiative levitation.
We find a depletion of [Fe/H] close to the TO ($T_{eff}\sim 6300$~K) of 
0.35~dex, compared to $\sim$0.30~dex from \citet{richard02} results. 
Around the TO ($T_{eff}\sim 6200$~K) of a [Fe/H]=$-$1.31, 
12~Gyr isochrone we find a [Fe/H] depletion by 
0.27~dex, compared to $\sim$0.25~dex from the corresponding \citet{richard02} isochrone.

\section{Discussion}

This section presents a detailed comparison between our Li 
abundance measurements and theoretical models, with the aim to assess the consistency 
between the initial A(Li) for M4  stars and the WMAP+BBNS value.

\subsection{Comparison with the theory: the role of the diffusion}

\begin{figure}
\includegraphics[width=84mm]{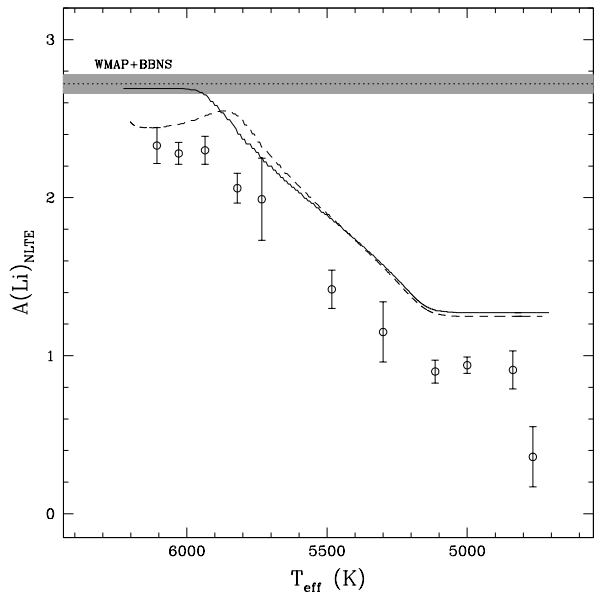}
\caption{
Behaviour of A(Li) as a function of $T_{eff}$ for our sample of stars 
grouped in eleven $T_{eff}$ bins. The curves indicate the theoretical predictions 
with (dashed line) and without (solid line) diffusion. 
Dotted line denote the primordial 
A(Li) computed by \citet{cyburt08} and the grey area the 
corresponding uncertainty.}
\label{fig7n}
\end{figure}

\begin{figure}
\includegraphics[width=84mm]{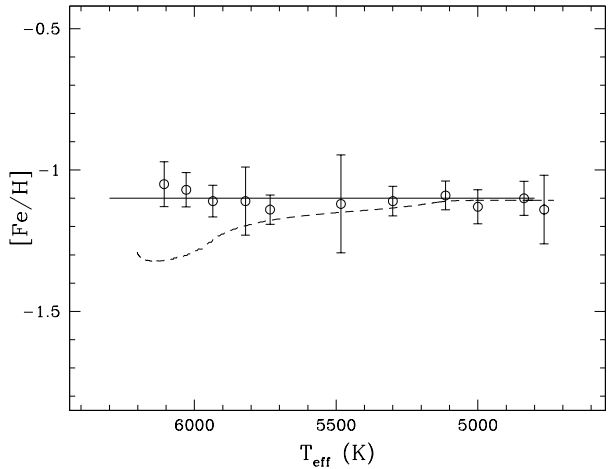}
\caption{
Behaviour of [Fe/H] as a function of $T_{eff}$ for our sample of stars 
grouped in eleven $T_{eff}$ bins (same binning of Fig.~\ref{fig7n}). 
The curves indicate the theoretical predictions 
with (dashed line) and without (solid line) diffusion.}
\label{fern}
\end{figure}

From the discussion in Sect.~4 it is clear that 
also M4 exhibits the discrepancy between the A(Li) value measured at the TO 
and the WMAP+BBNS estimate. 
We first investigate the role played by the atomic diffusion, that has 
been suggested by other authors as a possible solution to the problem, 
at least in case of NGC~6397 \citep{korn06}.
Fig.~\ref{fig7n} and \ref{fern} display the comparison between the measured surface 
Li and Fe abundances (grouped in 100--150~K wide temperature bins
only for sake of clarity 
\footnote{Note that the adoption of different binning steps 
does not change dramatically the results.})
and the predictions of evolutionary models with and 
without atomic diffusion, from the TO to the lowest temperature limit of our sample. 
For sake of comparison, we also plot 
the WMAP+BBNS A(Li) value used in the initial composition of our models, and the 
related uncertainty. The value A(Li) at the TO in the calculations without diffusion 
is very similar to the initial value (lower by only 0.03 dex) because the pre-MS Li 
depletion at these metallicities and for the TO mass of M4 is negligible.
This value of A(Li) stays constant (within the uncertainties)
from the TO to a point along the SGB with $T_{eff}\sim$5900-6000 K; 
below this temperature, the surface Li
abundance drops steadily, for the inward extending base of the convective 
envelope is reaching increasingly hotter layers where Li has been burned. 
This surface depletion ends when the convective envelope attains its 
maximum depth at the beginning of the RGB.

In the model with diffusion,
the surface Li around the TO is depleted by about 0.2--0.25~dex, in agreement with 
results by \citet[][see their Fig. 9]{richard02}. With decreasing temperature along 
the SGB, the surface A(Li) starts to rise again due to deepening of the convective
region, that returns to the surface some of the Li diffused outside the envelope 
along the MS. A(Li) reaches a peak when $T_{eff}\sim$5800--5900 K, with an 
abundance within $\sim$0.1~dex from the initial one. Moving to lower 
effective temperatures, the surface Li abundance 
starts to drop, because the convective envelope reaches regions where Li had
been burned due to the higher temperatures of the stellar matter. As for the case 
without diffusion, surface depletion ends when the convective envelope attains
its maximum inward extension.

\begin{figure}
\includegraphics[width=84mm]{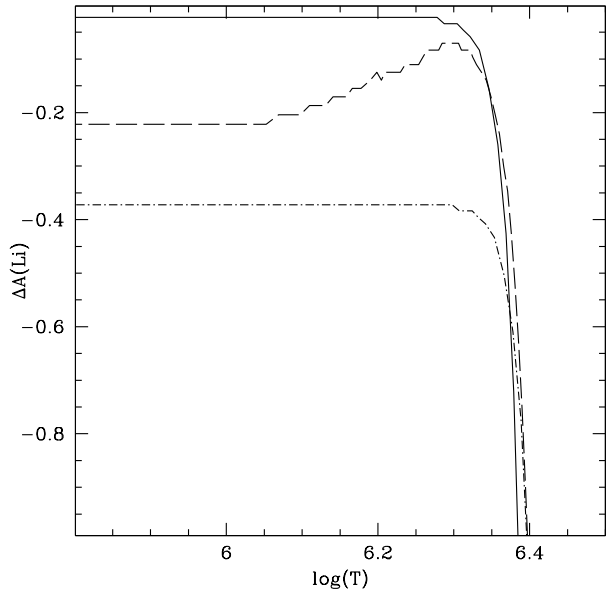}
\caption{Li abundance profile (in terms of the difference 
with respect to the initial value adopted in the calculations) as a function of temperature, 
for a model corresponding to M4 TO stars, including (dashed line) and neglecting 
(solid line) diffusion. The dot-dashed line denotes the abundance profile 
needed to reproduce the measurements of A(Li) in the cluster (see Fig.~\ref{fig6}).
}
\vspace{8mm}
\label{profiles}
\end{figure}

Fig.~\ref{profiles} helps to better clarify this behaviour. 
The Li abundance profile inside a model representing the cluster TO stars is displayed 
as a function of the temperature, for calculations with and without diffusion.
The model calculated without diffusion displays a constant A(Li) even beyond 
the convective envelope boundary, located at log($T$)$\sim$6.02, down to layers 
where the temperature is high enough to burn Li. The very small ($\sim$ 0.03~dex) difference 
with respect to the initial value is due to Li-burning during the pre-MS stage. 
When the model evolves along the SGB, convection deepens. As long as the bottom of 
the convective envelope stays within the temperature range where the Li profile is constant, 
the surface abundance does not change. When the base of the convective envelope reaches 
log($T$)=6.3-6.35, it engulfs Li-depleted layers and the surface A(Li) decreases steadily, 
until the envelope reaches its maximum extension at the end of the SGB phase. 
At this point the surface depletion ceases and the surface 
A(Li) predicted by the models stays constant during the following RGB evolution.
In case of models with diffusion, below the base of the convective envelope A(Li) first increases, 
due to the accumulation of Li diffused below the convection zone during the MS. When 
T reaches the Li-burning temperature, A(Li) drops fast, as in the model without diffusion. 
Along the SGB evolution, the base of the convective region engulfs first the layers 
with increased A(Li), and this causes the observed surface abundance increase after the TO. 
When convection goes deeper than log($T$)=6.3-6.35, the behaviour of the surface A(Li) becomes 
similar to the case without diffusion.
It is important to notice that the observed constant value of 
A(Li) along the RGB -- before the RGB bump mixing episode that is not accounted for by the models -- 
provides an additional constraint on the Li content at the TO, that is rarely considered 
in the majority of the chemical analyses.
In fact, given that the maximum depth of the convective envelope attained at the base of the RGB is independent 
of the exact value of the initial Li abundance, 
the predicted A(Li) along the lower RGB is determined by the internal Li abundance profile at the TO.

As for the trend of [Fe/H] with $T_{eff}$, 
the model without diffusion displays essentially a constant [Fe/H] along the whole 
evolution, equal to the initial value (the effect 
of the {\sl First Dredge-Up}, that decreases slightly the surface H-abundance,  
is negligible). Instead the model with diffusion has an 
abundance lower by 0.21~dex around the TO, due to the diffusion of Fe below 
the convective envelope during the MS phase. 
Along the SGB [Fe/H] increases 
steadily, for the increasing depth of the surface convective region brings back 
to the surface most of the Fe previously diffused from the convective envelope.
When the model reaches the RGB, the surface Fe abundance is only very 
slightly lower than the initial value, and stays constant thereafter. 
We estimate that the $T_{eff}$ of the TO stars should be 
increased by at least 300~K in order to lower the estimated iron abundance 
to values consistent  with the predictions of models including atomic diffusion, 
while keeping the same $T_{eff}$ for the RGB stars.
However, the resulting $T_{eff}$ for the TO stars would be incompatible 
with the $T_{eff}$ of theoretical isochrones, with or without diffusion.\\   

Some considerations can be drawn from the comparison between the observations and 
the theoretical models:\\ 
(1) we measure a uniform value of [Fe/H] along all the evolutionary sequences, that is  
reproduced by the model without diffusion. Taken at face value, this 
suggests that a process able to inhibit atomic diffusion from the convective 
envelope must be at work in the M4 stars;\\ 
(2) both the models with and without diffusion, that start from 
a WMAP+BBNS initial Li abundance, predict A(Li) for TO stars higher 
than our measurements in M4 (see Fig.~\ref{fig7n});\\ 
(3)  there is no observational evidence within the current observational errors
for the A(Li) peak before 
the SGB drop, predicted by the models with atomic diffusion (see Fig.~\ref{fig7n});\\ 
(4) both the theoretical curves in Fig.~\ref{fig7n} match reasonably well the value of 
$T_{eff}$ ($\sim$5900 K) where Li dilution due to the penetration 
of the convective envelope starts, and the temperature ($T_{eff}\sim$5200--5300 K) 
where the {\sl First Dredge-Up} ends. 
However, a clear offset in A(Li) is recognized. In fact, the models predict 
A(Li)$\sim$0.30~dex higher than that observed among the SGB and RGB stars.
This offset cannot be erased by adopting different $T_{eff}$ scales (se Sect. 6.2).
It is interesting to note that also in 
NGC 6397 \citep[][see their Fig. 1]{korn06}, the Li abundance 
among the RGB stars is lower than that predicted by the models tuned to fit the
TO and SGB data 
(R05) by $\sim$0.2~dex (despite this vertical offset has been not 
discussed by the authors).

\subsection{Comparison with the theory: the role of the turbulence}

A quite debated route to solve the discrepancy between the WMAP+BBNS Li 
abundance and the measurements along the {\sl Spite Plateau} is to assume that, 
besides atomic diffusion, a turbulent mixing process occurring below the convective 
region is at work, partially or totally counteracting the effect of diffusion from the 
envelope. 
The models developed by R05 
account for atomic diffusion (including radiative levitation) plus 
an {\sl ad hoc} parametrization of turbulent mixing that limits the settling 
of Li (and other metals), 
and at the same time does not transport matter into the convective envelope where 
Li has been burned by nuclear reactions. In particular,
the models computed with reference temperature log($T$)=6.0 
\footnote{This parametrization makes use of a 
turbulent diffusion coefficient that is 400 times larger than the diffusion 
coefficient of He at a reference temperature $T$, and varies with density 
($\rho$) as $\rho^{-3}$.}
allow \citet{korn06} to reproduce the observed TO and SGB A(Li) measurements 
in NGC~6397, starting from a primordial 
A(Li)=2.54 (lower than more recent estimates from WMAP+BBNS calculations)
and the small [Fe/H] difference ($\sim$0.17 dex) they measure between TO and RGB stars.

Let's consider the internal Li profile for the model calculated with diffusion, as that
displayed in Fig.~\ref{profiles}. Any turbulence that brings back into the convective 
envelope some of the Li accumulated around log($T$)=6.3 will increase the surface A(Li) 
and, at the same time, lower the Li abundance peak below the envelope. As a consequence, 
also the early increase of surface A(Li) 
during the SGB phase tends to be erased. 
This mechanism works well for NGC~6397, where 
A(Li) for TO stars is larger than the value predicted by atomic diffusion models, 
and the early increase along the SGB is also lower than expected from diffusion \citep{korn06}.
The same efficiency of turbulence (i.e. the same reference temperature log($T$)=6.0 in the 
parametrization by R05) however, 
cannot work for M4 and probably other clusters.
Fig.~\ref{figfin} highlights the problem. The 
solid line represents the value of A(Li)   
in TO stars, as a function of the metallicity, predicted by our models 
including atomic diffusion (without radiative levitation) 
for a reference age of 11-12~Gyr, and starting from the WMAP+BBNS value of A(Li). 
As discussed in Sect.~8, the lowest metallicity for which these 
Li predictions are adequate is [Fe/H]$\sim -$2.0 dex, since below this value  
radiative levitation affects appreciably the predicted surface Li abundances 
\citep{richard02}. 
When TO stars are located above the solid line (as in the case of 
NGC 6397), atomic diffusion during the MS is too efficient in depleting the Li surface
abundance and one can reproduce the observations only by including some form of 
additional turbulence that brings back to the surface some of the Li diffused out 
of the convective envelopes, as done by \citet{korn06}. 
For objects located below the solid line -- like M4 -- however, 
the inclusion of turbulent mixing with the same efficiency as in NGC~6397 
worsens the discrepancy, 
because it increases the surface A(Li) predicted by fully efficient diffusion.

This suggests that -- within the assumption of an initial A(Li) equal to the 
cosmological value --  for the three highest metallicity clusters in Fig.~\ref{figfin}, 
including M4, the turbulence has to reach deeper regions where Li burning occurs (i.e. 
one needs a higher reference temperature in R05 parametrization of turbulence),
and Li-depleted material is dragged to the surface, thus lowering A(Li) below the value 
expected from atomic diffusion.
The same turbulence could also account for the constant Fe abundance observed 
in our M4 star sample, feeding back to the envelope iron that diffuses below the 
convection zone during the MS phase.

The observed trend of A(Li) with $T_{eff}$ constrains the shape of 
the Li stratification in M4 TO stars. The value of A(Li) measured at the TO 
and its constancy (within the measurement errors) along the early SGB 
(without the Li peak typical of the 
Li profile in case of diffusion), plus the constraint given by the efficiency of the 
Li-burning  reactions require a profile like the one displayed as a dot-dashed line 
in Fig.~\ref{profiles}. We have calculated 
the SGB and RGB evolution starting from a TO model 
with this 'observationally constrained' Li profile and the abundances of the other 
elements as in the calculations without diffusion, in the assumption that turbulence 
has mixed back into the fully mixed  envelope the amount diffused below the convection 
boundary. The resulting surface Li evolution (Fig.~\ref{fig6})
reproduces very well the abundance pattern up to the RGB bump, including the A(Li) 
values along the early RGB. 

\subsection{Can we really detect the signature of diffusion?}

As discussed in Sect.~4, the observed distribution of A(Li) 
in TO stars is formally compatible with a constant Li value, within the measurement errors. 
In the previous analysis we have therefore considered a constant A(Li) for all TO stars.
The question we address here is the following: given the random errors on our estimates of 
A(Li), is it possible  to detect the trend of A(Li) with $T_{eff}$ predicted by models 
with fully efficient diffusion (dashed line in Fig.~\ref{fig7n}) ?
To this purpose, we have created a synthetic sample of A(Li) measurements, drawing randomly 
35 values of $T_{eff}$ from the Li-isochrone displayed in Fig.~\ref{fig7n}, using a flat 
probability distribution for $T_{eff}\geq 5900$~K. Gaussian errors equal to 
our average random observational errors for the TO stars have been then applied to this 
synthetic sample, and the a F-test as described in Sect.~4 has been performed. 
This procedure has been repeated 100 times, and for each of these 100 samples the F-test 
returned a probability below 90\% that a spread in A(Li) is present. 
This means that our measurements of A(Li) cannot rule out the existence of a trend consistent 
with predictions from models with fully efficient diffusion. If this were to be the case, 
the observations are matched for 
an initial A(Li) $\sim$0.20-0.25~dex lower than that the WMAP+BBNS value. \\ 
However, a similar test performed on the [Fe/H] values predicted by the models with diffusion 
for whole $T_{eff}$ range of our data,
shows that the F-test should be able to detect the expected spread of values, which is 
contrary to the result for our observed sample. 
We have verified that the predicted [Fe/H] depletion has to be reduced by at least 
$\sim$70\% in order to become undetectable by the F-test, given our random 
measurement errors.
This means that the efficiency of diffusion has to be strongly suppressed by come counteracting 
turbulence and that the curvature in the Li abundance as a function of $T_{eff}$ for TO stars 
has to be much lower than what predicted by fully efficient diffusion, if not completely vanishing.

\subsection{The initial Li content of M4 stars}

\begin{figure}
\includegraphics[width=85mm]{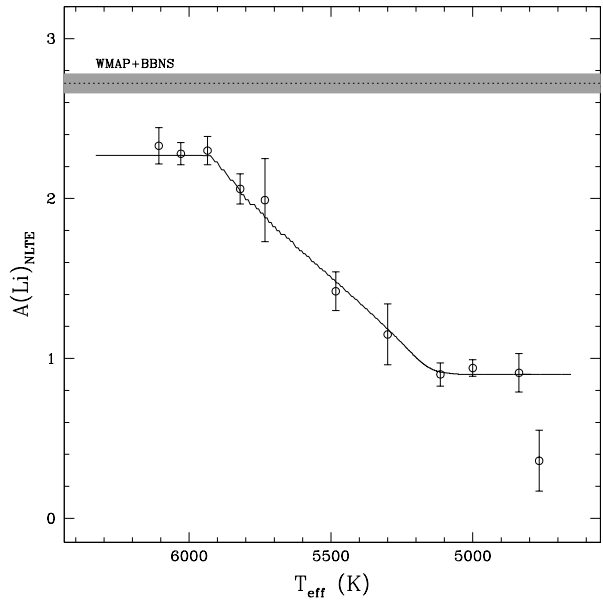}
\caption{Behaviour of A(Li) as a function of $T_{eff}$ for the stars of M4
(grouped in $T_{eff}$ bins as in Fig.~\ref{fig7n}), in comparison with the theoretical model 
computed by adopting the Li profile showed in Fig.~\ref{profiles} 
as dot-dashed line.}
\label{fig6}
\end{figure}

The discussion in the previous subsections and the fit displayed in Fig.~\ref{fig6} 
do not yet allow to identify univocally the pristine A(Li) in M4, because this value 
is strongly dependent on the details of the adopted turbulent mixing, that 
seems to play an essential part in the interpretation of the spectroscopic abundances.
Two {\sl extreme} situations can be envisaged to obtain the model plotted in Fig.~\ref{fig6}:\\ 
(i)~a pristine A(Li)=~2.72 dex and a turbulent mixing able to reach inner regions 
where Li is burned, thus dragging to the surface Li-poor material. This solution requires 
a higher reference temperature in R05 parametrization of turbulence
with respect to that proposed by \citet{korn06, korn06b}, that does not reach stellar 
regions where Li is depleted.
This solution allows to solve the discrepancy with WMAP+BBNS.\\ 
(ii)~a pristine A(Li)$\sim$2.35 dex and a turbulent mixing qualitatively similar to 
that of R05 with log($T$)=~6.09, showed in their Fig.~7. 
This efficiency would erase the Li peak at log($T$)=~6.3 (see Fig.~\ref{profiles}), 
by increasing the envelope A(Li) back towards the 
initial value. In this case another possibility to explain the discrepancy with the 
cosmological Li abundance -- apart from inadequacies 
of the BBNS standard model that could lead to an incorrect estimate of the 
cosmological A(Li) value \citep[see e.g.][]{jedam09} -- is to invoke that 
an amount of primordial 
Li could have been processed by Population III stars, before Population II 
stars were formed \citep{piau06}. The implication in this case is that Population II stars 
are not ideal to infer the cosmological Li abundance. While this could partially (up to 0.3 dex) 
solve the discrepancy between the primordial A(Li) and the initial 
abundance in M4, it also requires a complex and fine tuned scenario.
Even if Population III stars have depleted the Li abundance before the cluster formation, 
some form of turbulent mixing has still to be active below the convective envelope 
of M4 stars, to erase the A(Li) peak around log($T$)=6.3, that otherwise would produce 
an increase in the surface A(Li) at the beginning of the SGB evolution.\\

In light of these considerations, the solution proposed 
by \citet{korn06}, based on R05 parametrization of turbulence with reference temperature log($T$)=6.0 
and a initial Li abundance of 2.54 dex, may not be a {\sl universal} solution for the discrepancy 
with WMAP+BBNS calculations. This value of log($T$) adopted by \citet{korn06}
can account for the observations of NGC 6397, but would make the discrepancy worse for M4.
Instead, in case of M4 (and possibly also of NGC~6752 and 47~Tuc) one needs a turbulent mixing 
much deeper than in NGC~6397, that burns a sizable fraction of 
the initial Li.

\section{Conclusions}

We have measured the Li and Fe content for a large sample of stars along 
different evolutionary phases (TO, SGB and RGB) in the GC M4.
The main conclusion of this work can be summaryzed as follows:\\  
\begin{itemize} 
\item The Li content in the TO stars of M4 well resembles the Li abundance 
already observed in other GCs and in the Halo stars, thus confirming 
the discrepancy with the WMAP+BBNS predictions. \\
\item Qualitatively, the general behaviour of both Li and Fe with 
$T_{eff}$ is reproduced by models without diffusion, starting from 
the WMAP+BBNS A(Li) value but 
we recognize a systematic offset of $\sim$+0.3 dex 
between  the theoretical and the observed Li content along the SGB and RGB stars.\\
\item The behaviour of A(Li) 
as a function of $T_{eff}$ in TO/SGB stars is incompatible with the 
theoretical predictions of the models including atomic diffusion, 
starting from the WMAP+BBNS predicted value; This suggests that 
an additional turbulent mixing below the convective zone needs 
to be included in the models.\\ 
\item The general homogeneity and the value of A(Li) in the TO, early 
SGB and early RGB stars constrains to the shape of the Li stratification in M4 TO stars. 
A good match between theoretical models and the global behaviour of 
the Li abundance as a function of $T_{eff}$ can be obtained by assuming the  
WMAP+BBNS predicted value plus diffusion and a turbulent mixing  
reaching regions where Li is burned. 
The discrepancy with the WMAP+BBNS Li abundance cannot be 
solved by invoking the same 
reference temperature for the \citet{richard02} parametrization of turbulence, as 
proposed by \citet{korn06, korn06b} to explain the Li abundance measured 
in NGC 6397, because in that case the turbulent mixing does not reach layers  
where Li is burned.\\
\item Different (lower than the WMAP+BBNS predicted value) pristine A(Li) values 
for M4 are possible, depending on the adopted reference temperature for the turbulent 
mixing. This uncertainty is mainly due to our poor understanding 
of the true nature of the turbulent mixing that counteracts the atomic diffusion.
\\ 
\end{itemize}

The study of the Li content in M4 points out that the Li problem 
is far from being solved and new efforts are needed, both to enlarge the 
number of studied GCs and to refine theoretical stellar and 
cosmological models. 
Spectroscopic measurements of A(Li) below the TO would be crucial to put 
further constraints on the primordial Li in GC stars, given that moving 
down along the MS - even in case of fully efficient diffusion - one
expects to find Li abundances progressively closer to the primordial 
value. Moreover, a deeper understanding of the nature 
of the turbulent mixing occurring outside the convective region is needed to 
disentangle this problem.\\ 

We wish to thank Piercarlo Bonifacio and Luca Pasquini for the useful 
discussions and suggestions.
This research was supported 
by the Agenzia Spaziale Italiana (under contract ASI-INAF I/016/07/0), by the Istituto
Nazionale di Astrofisica (INAF, under contract PRIN-INAF2008) and by the Ministero
dell'Istruzione, dell'Universit\'a e della Ricerca.

\end{document}